# Energy spectrum of a 2D Dirac oscillator in the presence of a constant magnetic field and an antidot potential

Huseyin Akcay [1,a]   and  Ramazan Sever [2,b]

[1] Faculty of Engineering, Başkent University, Baglıca Campus, Ankara, Turkey

[2] Department of Physics, Faculty of Arts and Sciences, Middle East Technical University, 06531 Ankara, Turkey

## Abstract

We investigate the energy spectrum and the corresponding eigenfunctions of a 2D Dirac oscillator confined by an antidot potential in the presence of a magnetic field and Aharonov-Bohm flux field. Analytical solutions are obtained and compared with the results of the Schrödinger equation found in the literature. Further, the dependence of the spectrum on the magnetic quantum number and on the repulsive potential is discussed.

PACS: 03.65.G, 03.65.Pm, 73.63Kv

Keywords:  Two dimensional Dirac Oscillator, Constant magnetic field, Quantum dot and antidote

[a] E-mail: **akcay@baskent.edu.tr**

[b] E-mail: **sever@metu.edu.tr**

## 1. Introduction

During the last few years, the study of electrons in two dimensional systems have become an active research subject. The advances in nanofabrication technology has made it possible to confine two dimensional electrons. These nanostructures, such as, quantum dots, antidots, wells, wires have been a central point for both theoretical and experimental researchers. In condensed matter the studies of graphene is a major issue and the electrons in graphene may be viewed as massless changed fermions living in two dimensional space. Furthermore they show relativistic behavior. Currently, parabolic potentials [1] are often used to describe these confined electron systems. Despite its simplicity, it appears to be a good approximation for the quantum dot structure. An antidot [2-5] is created when a potential hill is introduced into a two dimensional electron system. These structures present an attractive tool to study quantum mechanics of interacting electrons [6-8]. The solutions of the Schrödinger equation for an electron in a homogeneous magnetic field with parabolic potential were obtained long ago [9,10]. Therefore, a large body of papers discuss the confinement of electrons in the framework of the Schrödinger equation. When the relativistic Dirac electrons with parabolic confinement are considered one needs to use numerical methods. The relativistic extension of a parabolic confining potential can be done using the Dirac oscillator [11]. Recently this approach has been used [12, 13] to study 2D Dirac electrons in the presence of magnetic field. In this paper we are studying 2D-Dirac oscillator with an additional antidot potential in presence of a constant magnetic field. The quantum antidot structure in the presence of repulsive and confining quadratic potentials has been studied in the framework of the Schrödinger equation [14]. There has been also some experimental studies on the thermodynamics (magnetization) and spectral properties of a two-dimensional electron gas with an antidot in magnetic field [15, 16].Therefore it is important to understand the behavior of such systems under various conditions. Here we are considering 2D- electron gas containing an antidot and modeling the antidot with a repulsive potential. A parabolic confining potential is provided by the oscillator term. This potential restricts the wave functions to a finite region if required.

İn the next section, we write the wave equation of the 2D Dirac oscillator in the presence of a constant magnetic field perpendicular to the plane where the electrons are confined to move. We add a parabolic confining potential and obtain analytical solutions of the wave equation. In section 3, we discuss the behavior of the energy levels and the wave functions. The nonrelativistic limit is also considered in this section.

## 2. Solution of the 2D Dirac equation

Time independent 2D Dirac equation with a scalar potential S, a vector potential V and a tensor potential U [1,18] can be written as

$$[c\vec{\alpha}.\vec{p} + V + \beta(m^*c^2 + S) + i\beta\vec{\alpha}.\hat{\rho}U]\psi = E\psi. \tag{1}$$

Where $\alpha^1 = \begin{pmatrix} 0 & 1 \\ 1 & 0 \end{pmatrix}$, $\alpha^2 = \begin{pmatrix} 0 & -i \\ i & 0 \end{pmatrix}$, $\alpha^3 = \begin{pmatrix} 1 & 0 \\ 0 & -1 \end{pmatrix}$ are the 2x2 Pauli matrices. We are going to use the polar coordinates ($\rho, \phi$) in this work. The tensor potential U, which is linear in $\rho$, is introduced using the substitution $\vec{p} \to \vec{p} - im^*\omega\beta.\vec{\rho}$. This replacement was proposed by Moshinsky and Szczepaniak [11]. It has been applied for several problems in hadron spectroscopy [19, 20], in quark confinement, in broken symmetries [21,22] and in semiconductor physics [23].in the investigation of the thermodynamic properties of some physical potentials [24], The interaction with the magnetic field is introduced by means of the minimal substitution $\vec{p} \to \vec{p} + (e/c)\vec{A}$. The system is assumed to be in a uniform magnetic field B and an Aharonov-Bohm (AB) flux field. Thus the vector potential will be a sum of two terms $\vec{A} = \vec{A}_1 + \vec{A}_2$ such that $\vec{\nabla}x\vec{A}_1 = \vec{B}$ and $\vec{\nabla}x\vec{A}_2 = 0$. Here $\vec{A}_2$ describes AB flux $\phi_{AB}$ created by the solenoid [14]. The nonzero components of the vector potentials are given by $A_{1\phi} = \frac{B}{2}\rho$, $A_{2\phi} = \frac{\phi_{AB}}{2\pi\rho}$ in the cylindrical coordinate system. First, we write the terms of Eq. (1) in terms of polar coordinates as

$$i\beta\vec{\alpha}.\hat{\rho}U = i\beta(\cos(\phi)\alpha^1 + \sin(\phi)\alpha^2)U = \begin{pmatrix} 0 & ie^{im\phi} \\ -ie^{-im\phi} & 0 \end{pmatrix} U,$$

$$e\vec{\alpha}.\vec{A} = \begin{pmatrix} 0 & -ie^{-i\phi}(eA_\phi) \\ ie^{i\phi}(eA_\phi) & 0 \end{pmatrix}, \quad c\vec{\alpha}.\vec{p} = (-i\hbar c)\begin{pmatrix} 0 & e^{-i\phi}(\partial_\rho - \frac{i}{\rho}\partial_\phi) \\ e^{i\phi}(\partial_\rho + \frac{i}{\rho}\partial_\phi) & 0 \end{pmatrix}.$$

For the two- component wave functions we use the following notation

$$\psi(\rho, \phi) = \frac{1}{\sqrt{2\rho}} \begin{bmatrix} F(\rho)e^{im\phi} \\ iG(\rho)e^{i(m+1)\phi} \end{bmatrix}. \tag{2}$$

We choose equal vector and scalar potentials. Our motivation for this choice is that, this makes it possible to introduce an antidote potential and obtain analytical bound state solutions [14, 25,26] of 2D- Dirac equation. Inserting these into Eq.(1) and using the fact that the potentials depend on the radial coordinate we find the following coupled system of first order differential equations for the radial wave functions

$$\hbar c[\frac{dF}{d\rho} - \frac{(m+\frac{1}{2})}{\rho}F] + [U - eA_\phi]F = -(E + m^*c^2 - \Delta)G \qquad (3a)$$

$$\hbar c[\frac{dG}{d\rho} + \frac{(m+\frac{1}{2})}{\rho}G] - [U - eA_\phi]G = (E - m^*c^2 - \Sigma)F \qquad (3b)$$

where $\Delta = V - S$ and $\Sigma = V + S$. As stated before we have chosen equal vector and scalar potentials thus $\Delta$ will be replaced with zero and $\Sigma$ with 2V. We parameterize the potentials as $V = \delta/(2\rho^2)$ and $U = (1/2)m^*c\omega_0\rho$. Using the expression for G obtained from Eq. (3a) and inserting it into Eq. (3b) we find the following second order differential equation

$$\frac{d^2F}{d\rho^2} + \frac{1}{\rho^2}[-(m+\alpha)^2 + \frac{1}{4} - \frac{(E+m^*c^2)}{(\hbar c)^2}\delta]F - \rho^2[\frac{m^*}{2\hbar}\omega_c + \lambda)]^2 F$$
$$- (m+\alpha+1)(\frac{m^*}{\hbar}\omega_c + 2\lambda)F = -\frac{(E^2 - m^{*2}c^2)}{(\hbar c)^2}F \qquad (4)$$

Where we have used the definitions $\alpha = \frac{e\phi_{AB}}{2\pi\hbar c}$, $\omega_c = \frac{eB}{m^*c}$ and $\lambda = \frac{m^*}{\hbar}\omega_0$. In order to simplify the notation we will first define some new parameters and then rewrite Eq.(4) in terms of these parameters. We define

$$\Lambda_1 = \frac{1}{4}(\frac{m^*}{2\hbar}\omega_c + \lambda)^2, \qquad \Lambda_2 = \frac{(E^2 - m^{*2}c^4)}{4(\hbar c)^2} - \frac{1}{4}(m+\alpha)(\frac{m^*}{\hbar}\omega_c + 2\lambda), \qquad (5)$$

$$\Lambda_3 = \frac{1}{4}(m+\alpha)^2 - \frac{1}{16} + \frac{(E+m^*c^2)}{(2\hbar c)^2}\delta .$$

and using these definitions we write Eq.(4) as follows

$$\frac{d^2F}{d\rho^2} - 4\Lambda_1\rho^2 F - 4\Lambda_3\frac{1}{\rho^2}F + 4\Lambda_2 F = 0 . \qquad (6)$$

Furthermore, it is more convenient to introduce a new variable $s = \rho^2$ and write the last equation in terms of s which gives

$$\frac{d^2F}{ds^2} + \frac{1}{2s}F + \frac{1}{s^2}(-\Lambda_1 s^2 + \Lambda_2 s - \Lambda_3)F = 0 . \qquad (7)$$

This differential equation is the hypergeometric type and thus it has polynomial solutions [27]. First, we consider the asymptotic form of this equation and try a solution of the form

$$F(s) = \exp(-ps)s^q y(s) \tag{8}$$

Where p and q are some constants and $y$ is a new function to be determined. Inserting Eq.(8) into Eq.(7) leads to a second order differential equation for $y$. When we write this differential equation for $y$ in terms of a variable $z = 2ps$ we find

$$z\frac{d^2 y}{dz^2} + (k+1-z)\frac{dy}{dz} + ny = 0. \tag{9}$$

Where k =2q-1/2, and n is an integer. We have studied differential equations of the parametric form given by Eq. (7) using the polynomial method and the details of this derivation can be found in Ref. (28). Eq.(9 is the Laguerre's associated equation [27] and its solutions are the associated Laguerre polynomials $L_n^k(z)$. In the derivation of Eq. (9) we have used the fact that $F$ is a bound state wave function and should satisfy the normalization conditions. These boundary conditions produced the integer n and also lead to the following relations between the parameters [28, 29]

$$q = \frac{1}{4}(1+\gamma), \gamma = \sqrt{1+16\Lambda_3} = \sqrt{1+4(m+\alpha)^2 + 4\frac{(E+m^*c^2)}{(\hbar c)^2}\delta}, \ k = \frac{\gamma}{2} \tag{10a}$$

$$p = \sqrt{\Lambda_1} = \frac{1}{4}\frac{m^*}{\hbar}\omega, \ \omega = \omega_c + 2\omega_0, \ \frac{1}{2}p + 2qp + 2np - \Lambda_2 = 0. \tag{10b}$$

The derivations of these relations are not complicated and we do not want to repeat them here.

## 3. The energy levels and the wave functions

We start with Eq. (10b) which leads to a formula for the possible energy values. In fact, inserting q and p from Eq. (10a) into Eq. (10b) we find $\sqrt{\Lambda_1}(2n+1+\frac{1}{2}\gamma) = \Lambda_2$.
The parameters $\Lambda_1$ and $\Lambda_2$ are defined in Eq. (5) and replacing these parameters in the last equation gives

$$\frac{E^2 - m^{*2}c^4}{(\hbar c)^2} = \frac{m^*}{\hbar}|\omega|(2n+1+\sqrt{(m+\alpha)^2 + \frac{E+m^*c^2}{(\hbar c)^2}\delta}) + \frac{m^*}{\hbar}(\omega)(m+\alpha). \tag{11}$$

Now let us look at the nonrelativistic limit for these energy levels. For this we define $\varepsilon = E - m^*c^2$ and use the substitution $(E+m^*c^2) \to 2m^*c^2$ in Eq. (11) and find

$$\varepsilon_{nm} = \hbar|\omega|(n+\frac{1}{2}+\frac{1}{2}\sqrt{(m+\alpha)^2+b^2}\,)+\frac{\hbar}{2}\omega(m+\alpha). \tag{12}$$

Where $b^2 = \dfrac{2m^*}{\hbar^2}\delta$ which is a dimensionless constant. In the absence of the tensor, and vector potentials ($\omega_0 = 0, b = 0$) and for $\alpha = 0$, Eq. (12) reduces to the following formula

$$\varepsilon_{nm} = \hbar\omega_c(n+\frac{|m|+m+1}{2}) \tag{13}$$

These are the well-known Landau levels [30]. In Fig. (1) the energy levels $\varepsilon_{0m}$, given by Eq. (12), are plotted (in $\hbar\omega$ units) as a function of magnetic quantum number $m$. We observe that these energy levels (denoted by squares) are degenerate for negative values of $m$. When $\alpha$ is not zero we get the shifted curve (indicated with bars) which is also degenerate. This degeneracy is removed when the antidot potential is present (represented by cross signs). For plotting these we have used the numerical values $\alpha = 8$ and $b = 10$. To obtain these we have fixed the values of $B, \delta$ accordingly [14] and assumed that $\omega_0$ is near to $\omega_c$. Going back to Eq. (11) we use numerical methods for the calculation of the energy levels. For this calculation, it is more practical to study energy levels in units of $m^*c^2$. We define $\chi = E/m^*c^2$ and write Eq. (11) in terms of this variable. It gives

$$(\chi^2 - 1) = \kappa\{(n+\frac{1}{2}+\frac{1}{2}\sqrt{(m+\alpha)^2+(\chi+1)\frac{b^2}{2}}\,)+\frac{(m+\alpha)}{2}\}\,. \tag{14}$$

Where $\kappa$ is dimensionless constant which is defined as $\kappa = \dfrac{4\hbar\omega}{m^*c^2}$. In Fig. (2) we plot $\eta$ with respect to m. This quantity is defined as $\eta = (\chi^2-1)(m^*c^2)/(2\hbar\omega)$ and reduces to $\varepsilon/\hbar\omega$ in the nonrelativistic limit. Note that it is expressed in $\hbar\omega$ units. In the figure squares corresponds to the energy levels in the absence of the dot potential (the repulsive potential). The upper curve corresponds to the energy levels in the presence of all the potentials. From the figure we see that the antidot potential removes the degeneracy also for the relativistic expression. We note that for $\alpha = 0$ and $\omega_c = 0$ our formula reduces to the Fock-Darwin levels [31]. A similar graphical analysis can be given for this case.

The wave functions are given by Eq. (8) and using the definitions given for q and p we write

$$F_{nm}(\rho) = N\exp(-\frac{m^*}{2\hbar}\omega\rho^2)\rho^{\frac{\gamma}{2}}L_n^{\gamma}(\frac{m^*}{2\hbar}\omega\rho^2), \quad N = \{\frac{(2p)^{\frac{\gamma}{2}+1}}{\sqrt{\pi}}(\frac{n!}{(n+\frac{\gamma}{2})!})\}^{\frac{1}{2}} \quad . \tag{15}$$

Here N is the normalization constant. The probability density P obtained with Eq. (15) is plotted with respect to the distance is plotted in Fig. (3).Eq. (15) The figure show the effect of the repulsive potential and AB field. We observe that the probability density of the electron is nearly zero in the presence of the antidot potential and AB field for small values of $\rho$. The antidot potential extends the zero regions towards the right hand side. The Dirac oscillator has an effect in the opposite direction and can be arranged to make confinement more effective.

## 4 Conclusions

The 2D Dirac oscillator is investigated in the presence of a constant magnetic field and a repulsive potential. The repulsive potential is constructed with a proper choice of the vector and scalar potential s. The radial wave equations are solved analytically and the energy spectrum and the wave function are obtained. The behavior of the energy spectrum and the wave function are investigated. The degeneracy of the energy spectrum with respect to the magnetic quantum number and its removal by the repulsive potential term is discussed. The nonrelativistic limit is also considered and our results are compared with results of the Schrödinger theory which already exist in literature.

**FİGURE CAPTİONS :**

Fig .(1). The squares represent the Landau levels in the presence of a strong magnetic field .The curve with the cross signs represents the spectrum when all potentials are present. While the curve with the dots is the Landau spectrum in strong magnetic field and a AB flux field.

Fig.(2). The open circles indicate the levels in the presence of the magnetic field and the antidote potential. The other curve represents the Landau levels ın the strong magnetic field

Fig. (3a). The probability density in the absence of the repulsive potential.

Fig. (3b).The density function in the presence of the antidote potential

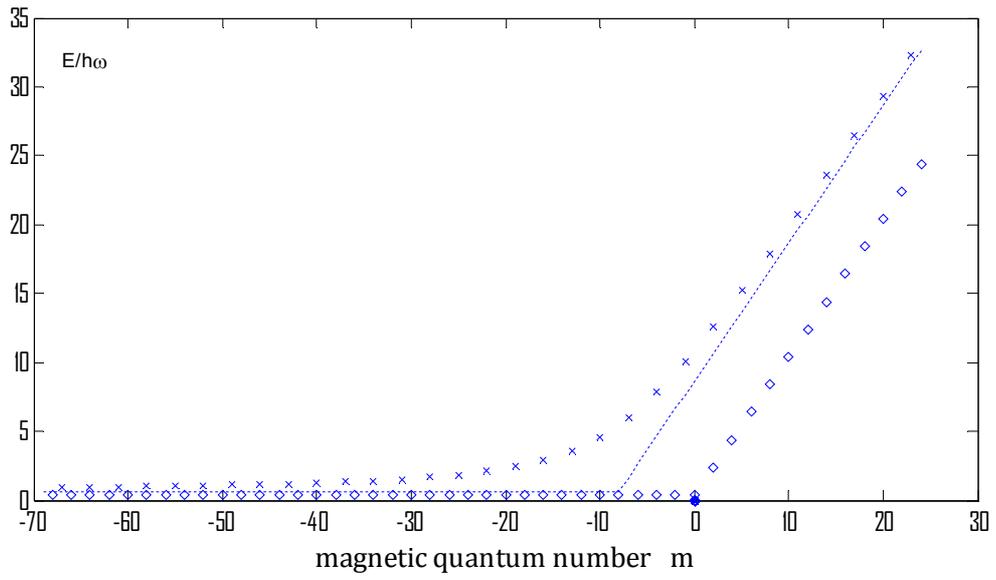

Figure-1

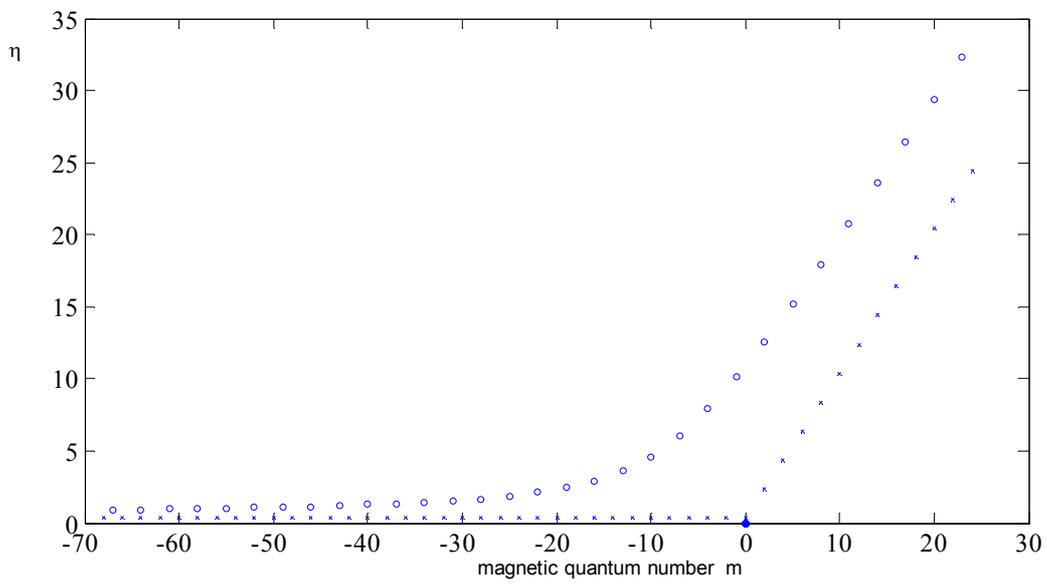

Figure-2

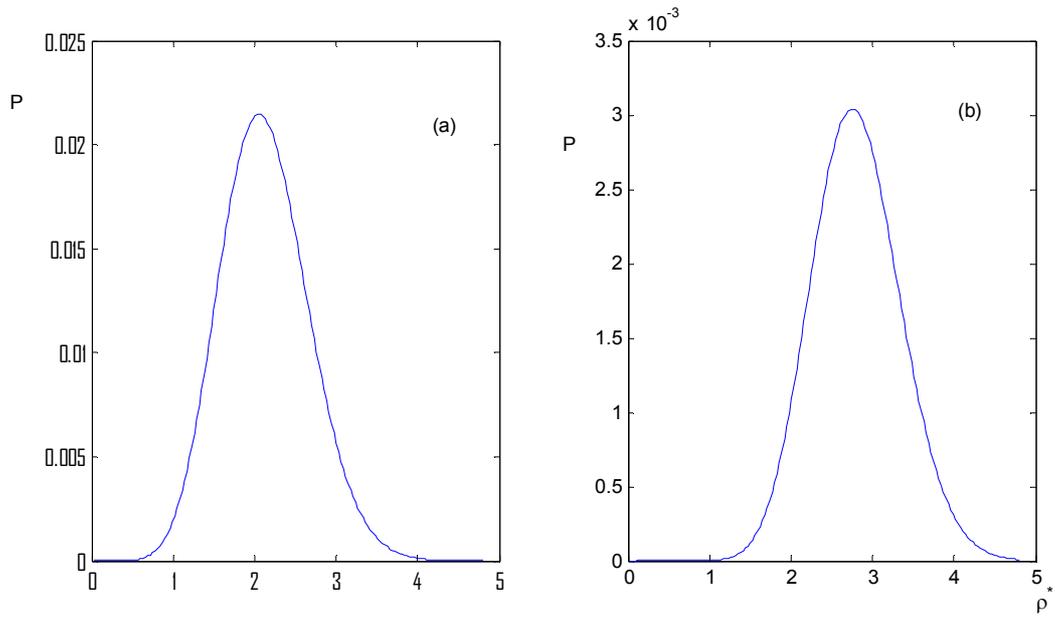

Figure-3